\documentclass[aip,apl,amsmath,amssymb,reprint]{revtex4-1}

\usepackage{graphicx}
\usepackage{dcolumn}
\usepackage{bm}
\usepackage{microtype}
\usepackage[utf8]{inputenc}
\usepackage[T1]{fontenc}
\usepackage{mathptmx}

\newcommand{\CC}{\text{C}_\text{B}\text{C}_\text{N}}  

\begin{document}

\title{Carbon dimer defect as a source of the 4.1 eV luminescence in hexagonal boron nitride}

\author{M. Mackoit-Sinkevičienė}
\email{mazena.mackoit@ftmc.lt}
\affiliation{Center for Physical Sciences and Technology (FTMC), Vilnius LT-10257, Lithuania}

\author{M. Maciaszek}
\email{marek.maciaszek@pw.edu.pl}
\affiliation{ Faculty of Physics, Warsaw University of Technology, Koszykowa 75, 00-662 Warsaw, Poland}

\author{C. G. Van de Walle}
\affiliation{Materials Department, University of California, Santa Barbara, California 93106-5050, USA}

\author{A. Alkauskas}
\affiliation{Center for Physical Sciences and Technology (FTMC), Vilnius LT-10257, Lithuania}
\affiliation{Department of Physics, Kaunas University of Technology, Kaunas LT-51368, Lithuania}

\date{\today}

\begin{abstract}
We propose that the carbon dimer defect {$\CC$} in hexagonal boron nitride gives rise to the ubiquitous narrow luminescence band with a zero-phonon line of 4.08 eV (usually labeled the 4.1 eV band). Our first-principles calculations are based on hybrid density functionals that provide a reliable description of wide band-gap materials. The calculated zero-phonon line energy of 4.3 eV is close to the experimental value, and the deduced Huang-Rhys factor of $S \approx 2.0$, indicating modest electron-phonon coupling, falls within the experimental range. The optical transition occurs between two localized $\pi$-type defects states, with a very short radiative lifetime of 1.2 nanoseconds, in very good accord with experiments.
\end{abstract}

\maketitle

Layered materials with inter-layer van der Waals (vdW) bonding have recently attracted a lot of interest due to their distinct chemical and physical properties \cite{Duong2017}. Among this class of systems hexagonal boron nitride (hBN) stands out because of its large band gap of 6.08 eV \cite{Cassabois2016}. Advances in growth techniques have improved the materials quality \cite{Watanabe2004} and enabled the growth of single-layer hBN, opening up applications in electronic and optoelectronic devices. Currently hBN is mainly used in a passive role, for example as a substrate and/or insulating layer in electronic devices made of graphene and other vdW materials \cite{Duong2017}, or as a dielectric for photonic crystal cavities \cite{Kim2018}. However, hBN can also be used as an {\it active} optoelectronic material, e.g., as an electron-pumped ultraviolet (UV) laser \cite{Watanabe2004}. The recent discovery \cite{Tran2016} that hBN can host bright and stable single-photon emitters in the visible spectral range has sparked huge interest in the application of hBN as a light source in quantum optics applications.

Photoluminescence (PL), cathodoluminescence (CL), and electroluminescence experiments dating back to the 1950s \cite{Larach1956} already revealed a strong emission band between 3.3 and 4.1 eV in bulk hBN. This near-UV emission was so prevalent in some early samples that the band gap of hBN was sometimes erroneously assumed to be just above 4 eV \cite{Solozhenko2001}. However, more careful spectroscopic experiments \cite{Era1981,Museur2008,Du2015} on better-quality material revealed that the 4 eV luminescence is defect-related and is composed of at least two bands with very distinct properties. One is a broad featureless band centered around 3.9 eV \cite{Museur2008}. The other is a much narrower band with a clearly distinguishable zero-phonon line (ZPL) at 4.08 eV (typically called the 4.1 eV band in the literature) and which is accompanied by a few phonon replicas \cite{Era1981,Museur2008,Du2015}. The dimensionless Huang-Rhys parameter, which quantifies electron-phonon coupling during optical transitions \cite{Stoneham}, was estimated to fall in the range $S=1-2$ for this band \cite{Vokhmintsev2019}. The PL of this structured band appears when excitation energies exceed the ZPL of 4.1 eV. At variance, the broad 3.9 eV band appears only at excitation energies larger than 5.0 eV \cite{Museur2008}. Furthermore, time-dependent luminescence associated with these bands possesses very distinct characteristics. The structured narrow band shows very fast single-exponential decay with a lifetime $\tau=1.1-1.2$ ns \cite{Era1981,Museur2008}, while the wide band exhibits multi-exponential dynamics with the slowest components having decay times of a few 100 ns \cite{Museur2008}. All these results indicate a very distinct origin of the two bands, and from now on we will only discuss the structured 4.1 eV band.

Recently, single-photon emission associated with the 4.1 eV band has been reported \cite{Bourrellier2016}. Fast electrons in a transmission electron microscope \cite{Mueret2015} were used to excite luminescence at $T=150$ K. Measurement of the second-order correlation function confirmed that photons originate at a single emitter. The lineshape and the lifetime \cite{Mueret2015} of the CL band were identical to those in ensemble measurements, confirming that in both experiments luminescence was caused by the same defect. These experiments have renewed the interest in the 4.1 eV band due to its potential use in quantum optics.

Despite the ubiquity of the 4.1 eV line, the microscopic nature of the defect that causes the luminescence is still not known. The intensity of the band increases drastically in both bulk crystals \cite{Era1981} and epitaxial layers \cite{Uddin2017} when carbon is purposely introduced during growth. Therefore, the involvement of carbon has been naturally assumed \cite{Era1981,Du2015}. It has been suggested \cite{Katzir1975,Du2015,Uddin2017} that the 4.1 eV emission is caused by a transition from either a shallow donor (a so-called donor-acceptor-pair or DAP transition) or from the conduction band (free-to-bound transition) to the neutral carbon acceptor on the nitrogen site, C$_{\text{N}}$. There are strong arguments against these scenarios. First, the time dynamics of DAP and free-to-bound transitions are inconsistent with the measured lifetime of $\tau = 1.1-1.2$ ns. For DAP transitions, the variation in donor-acceptor pair distances usually leads to marked non-exponential decay dynamics with very long tails \cite{Museur2008}, at odds with the single-exponential decay of the 4.1 eV line \cite{Era1981,Museur2008}. Regarding radiative free-to-bound transitions, these occur on a millisecond time scale at typical excitation conditions (carrier densities $\sim$$10^{17}$ cm$^{3}$) \cite{Stoneham}, significantly slower than the dynamics of the 4.1 eV line. An additional argument comes from our recent first-principles study \cite{Weston2018}, where the acceptor level of C$_{\text{N}}$ was found to be at 2.9 eV above the valence band maximum (VBM). Since the band gap of hBN is $\sim$6.1 eV, DAP and free-to-bound transitions should therefore have energies smaller than 3.2 eV, i.e., they should not appear in the UV at all. The fast nanosecond radiative decay dynamics of the 4.1 eV line \cite{Era1981,Mueret2015,Museur2008} indicates that this is a transition where the ground state and excited state are localized in close proximity, likely on the same defect. Recently, Korona and Chojecki \cite{Korona2019} used quantum chemistry calculations to suggest that carbon clusters made from two to four atoms give luminescence in the range from 3.9 to 4.8 eV in monolayer hBN. However, different UV lines were not discriminated in that study, and neither the stability of clusters nor parameters (lifetime and electron-phonon coupling) of optical transitions were investigated.

In this Letter we use first-principles density functional theory to show that the {$\CC$} complex, in which two carbon atoms substitute on nearest-neighbor sites in the hBN lattice, accounts for all known experimental facts about the 4.1 eV luminescence: the involvement of carbon, the energy of the transition, the very short radiative lifetime, and moderate electron-phonon coupling.

Our calculations are based on the hybrid density functional of Heyd, Scuseria, and Ernzerhof \cite{HSE}. In this approach, a fraction $\alpha$ of screened Fock exchange is admixed to the short-range exchange potential described by the generalized gradient approximation of Perdew, Burke and Ernzerhof \cite{PBE}. We use $\alpha=0.40$, for which calculations yield a band gap of 6.42~eV, consistent with the experimental gap\cite{Cassabois2016} when zero-point renormalization due to electron-phonon interactions~\cite{Antonius2015} is accounted for.
We used the projector-augmented wave approach \cite{PAW} with a plane-wave energy cutoff of 500 eV, and van der Waals interactions were included via the Grimme D3 empirical correction scheme \cite{Grimme2006}.  With these settings the calculated lattice parameters of hBN ($a = 2.49$ {\AA} and $c = 6.51$ \AA) and the enthalpy of formation for hBN (2.96 eV per formula unit) are in very good agreement with experimental values \cite{Tomaszkiewicz2002}. Defect calculations have been performed in orthorhombic supercells containing 240 atoms \cite{Weston2018} and with lattice vectors $5({\bf a}+{\bf b})$, $3({\bf a}-{\bf b})$, $2{\bf c}$, where ${\bf a}$, ${\bf b}$, ${\bf c}$ are vectors of the primitive hBN lattice. The Brillouin zone was sampled at the $\Gamma$ point. Ionic relaxation was carried out until Hellman-Feynman forces were less than 0.005 eV/\AA. Calculations have been performed using the Vienna Ab-initio Simulation package (\textsc{vasp}) \cite{vasp}.

\begin{figure}
\includegraphics[width=8.5cm]{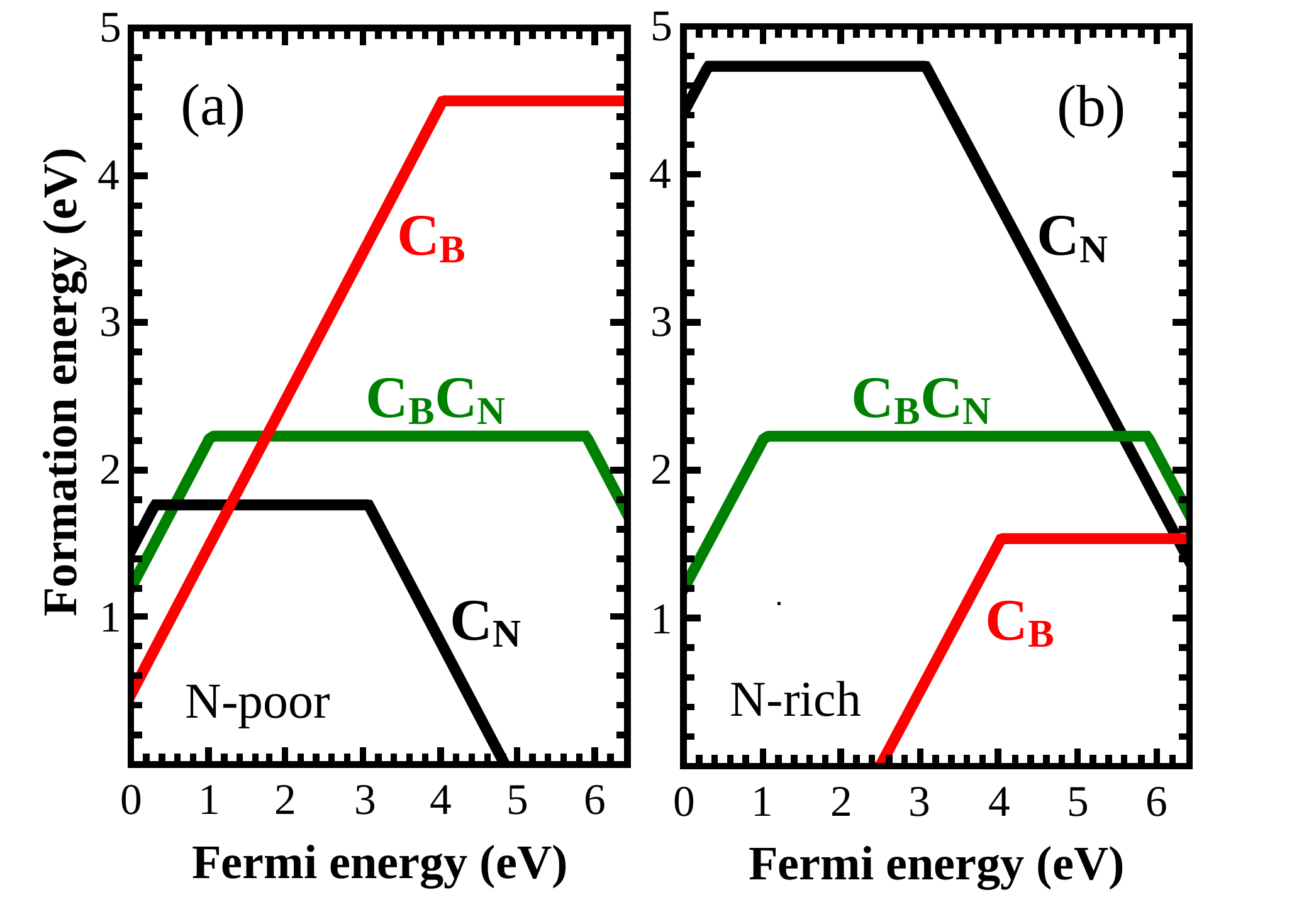}
\caption{Calculated formation energies vs the Fermi level for {$\CC$}, $\text{C}_\text{B}$, and $\text{C}_\text{N}$ defects under (a) N-poor and (b) N-rich conditions.}
\label{fig:formation}
\end{figure}

We start by calculating the formation energy \cite{Freysoldt2014} of the carbon dimer $E_{f}(\CC)$, which is given by:
\begin{eqnarray}
E_{f}(\text{C}_{\text{B}}\text{C}_{\text{N}})&=&
E_{\text{tot}}(\text{C}_{\text{B}}\text{C}_{\text{N}})-E_{\text{tot}}(\text{BN})+\mu_{\text{B}}+\mu_{\text{N}}
\nonumber
\\
&& -2\mu_\text{C} + q(E_{F}+E_V) + \Delta_q,
\label{eq:formation}
\end{eqnarray}
where $E_{\text{tot}}(\text{C}_\text{B}\text{C}_\text{N})$ is the total energy of the supercell containing one dimer, and $E_{\text{tot}}(\text{BN})$ is the total energy of a pristine supercell.
$\mu_{\text{N}}$ and $\mu_{\text{B}}$ are chemical potentials of nitrogen and boron; $\mu_{\text{N}}+\mu_{\text{B}}=\mu_{\text{BN}}=E_{\text{BN}}$, where $E_{\text{BN}}$ is the total energy of bulk BN per formula unit. $\mu_\text{C}$ is the chemical potential of carbon, set to the per-atom energy of a diamond crystal. In Eq.~(\ref{eq:formation}) $q$ is the charge of the defect, and $E_{F}$ is the Fermi level, referenced to the VBM $E_{V}$.  $\Delta_q$ is a finite-size electrostatic correction term \cite{Freysoldt2009}. We note that the formation energy of the dimer does not depend on individual chemical potentials $\mu_{\text{N}}$ and $\mu_{\text{B}}$, as $\mu_{\text{N}}+\mu_{\text{B}}=\mu_{\text{BN}}$. The calculated formation energy is shown in Fig.~\ref{fig:formation}, together with formation energies of $\text{C}_\text{B}$ and $\text{C}_\text{N}$ defects \cite{Weston2018}. For these two latter defects formation energies {\it do} depend on the chemical potentials of boron and nitrogen; only two limit cases are shown in Fig.~\ref{fig:formation}. For N-rich conditions $\mu_{\text{N}}^{\text{rich}}=1/2 E_{\text{tot}}(\text{N}_2)$, half the energy of the N$_2$ molecule; for B-rich (N-poor) conditions $\mu_{\text{B}}^{\text{rich}}=E_{\text{tot}}(\text{B})$, the energy of the B atom in elemental boron.

We find that {$\CC$} has three possible charge states, $q=-1$, $q=0$, and $q=+1$ (Fig.~\ref{fig:formation}). The neutral charge state is the most stable one throughout most of the band gap, with a formation energy of 2.2 eV. As can be seen in Fig.~\ref{fig:formation}, the formation energy of the dimer is not lower than those of simple substitutional defects for the two limiting cases of atomic chemical potentials. However, there is a wide range of chemical potentials ($\mu_{\text{N}}^{\text{rich}}-2.5 \text{ eV}<\mu_{\text{N}}<\mu_{\text{N}}^{\text{rich}}-0.7 \text{ eV}$) and Fermi levels for which {$\CC$} is more stable than either $\text{C}_\text{B}$ or $\text{C}_\text{N}$. In addition, if both $\text{C}_{\text{B}}$ and $\text{C}_{\text{N}}$ are present in the material (e.g., as a result of non-equilibrium growth), the formation of {$\CC$} is expected. For example, when $\text{C}_\text{B}^+$ binds to $\text{C}_\text{N}^-$ to form ($\CC$)$^0$, an energy of 3.1 eV is released, indicating an exothermic reaction. We conclude that whenever carbon is present during the growth of hBN, {$\CC$} should be a common defect.

\begin{figure}
\includegraphics [width=8.5cm]{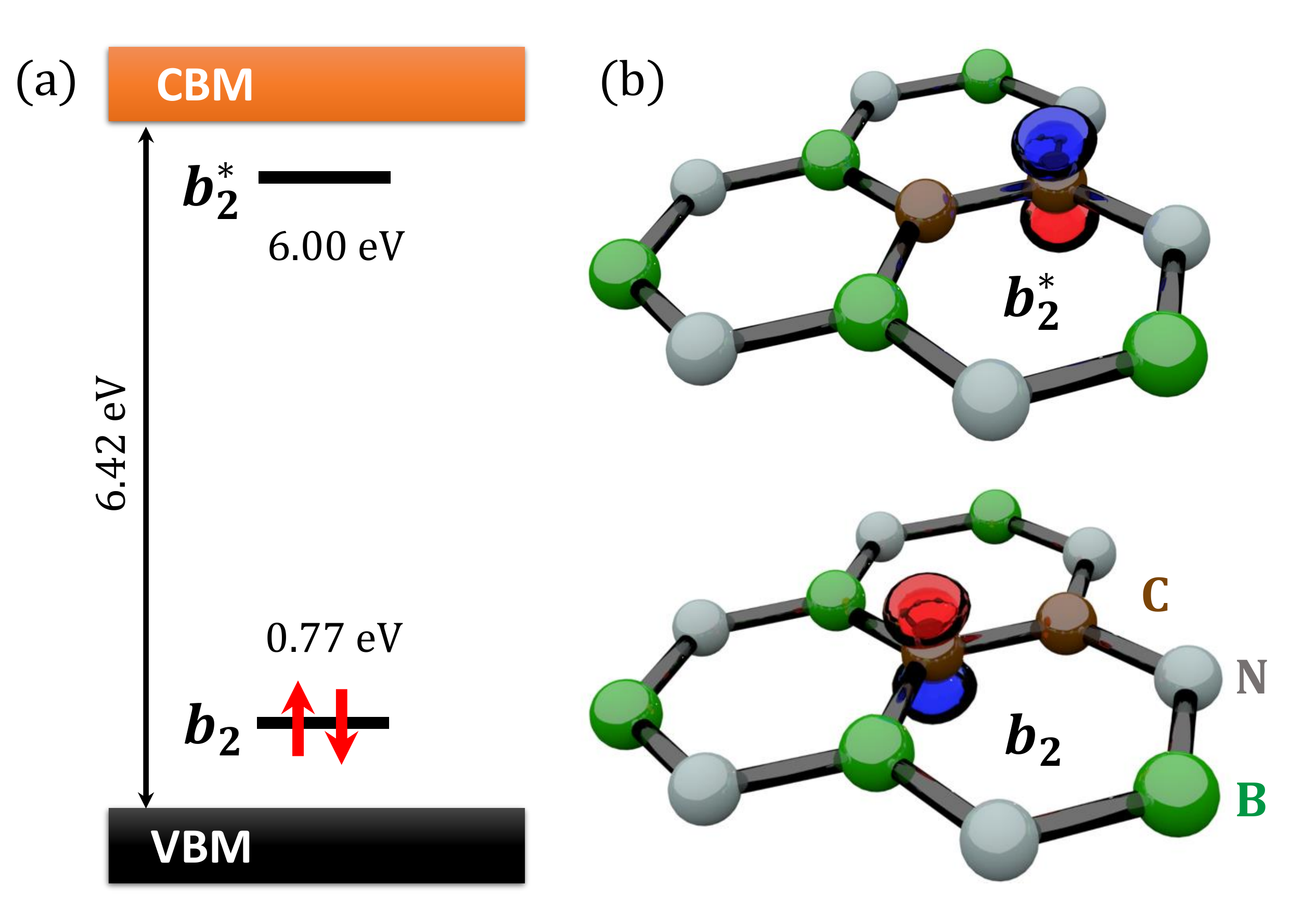}
\caption{
(a) Energies of Kohn-Sham states and (b) wave functions of the defect states of the neutral {$\CC$} complex in hBN.}
\label{fig:KS}
\end{figure}

We now turn to electronic properties of the dimer. In the neutral state, which is the one we will consider here, the dimer is non-magnetic ($S=0$). The Kohn-Sham electronic state diagram [Fig.~\ref{fig:KS}(a)] shows that there are two defect states in the band gap. The lower-lying state is a $p_z$ orbital localized on the ``acceptor'' site $\text{C}_{\text{N}}$, while the higher-lying state is a $p_z$ orbital on the ``donor'' site $\text{C}_{\text{B}}$ [Fig.~\ref{fig:KS}(b)]. The defect geometry belongs to the ${C_{2v}}$ point group, and both states can be labeled according to the irreducible representation $b_2$. To distinguish the two states, we label the upper one $b_2^*$.

In the ground state of the neutral dimer, the $b_2$ state is doubly occupied, while the $b_2^*$ state is empty, resulting in the electronic wave function $\vert b_{2}\bar{b}_{2}\rangle$ (symbols without a bar are for spin-up electrons, symbols with a bar for spin-down). This is a singlet state $^1A_1$. In the ground state the length of the C$-$C, C$-$N, and C$-$B bonds are 1.361, 1.391, and 1.497 {\AA}, respectively (cf.~the nearest-neighbor distance of 1.435 {\AA} in bulk hBN). The excited state is obtained when one $b_2$ electron is promoted to the $b_2^*$ state, yielding the wave function  $\frac{1}{\sqrt{2}}(\vert b_{2}\bar{b}^*_{2}\rangle - \vert b_{2}^*\bar{b}_{2}\rangle )$, also a $^1A_1$ state. We calculate the energy and the resulting geometry of the defect in the excited state using the so-called delta self-consistent field approach (${\Delta}$SCF) \cite{Jones1989} with constrained orbital occupations, as explained in the Supplementary Material. In the excited state there is a slight geometry rearrangement: the C$-$C bond elongates by 7\% to 1.456 {\AA}, while C$-$N and C$-$B bond lengths change by less than 1.5\% (to 1.372 and 1.499 {\AA}, respectively).

\begin{figure}
\includegraphics [width=8.5cm]{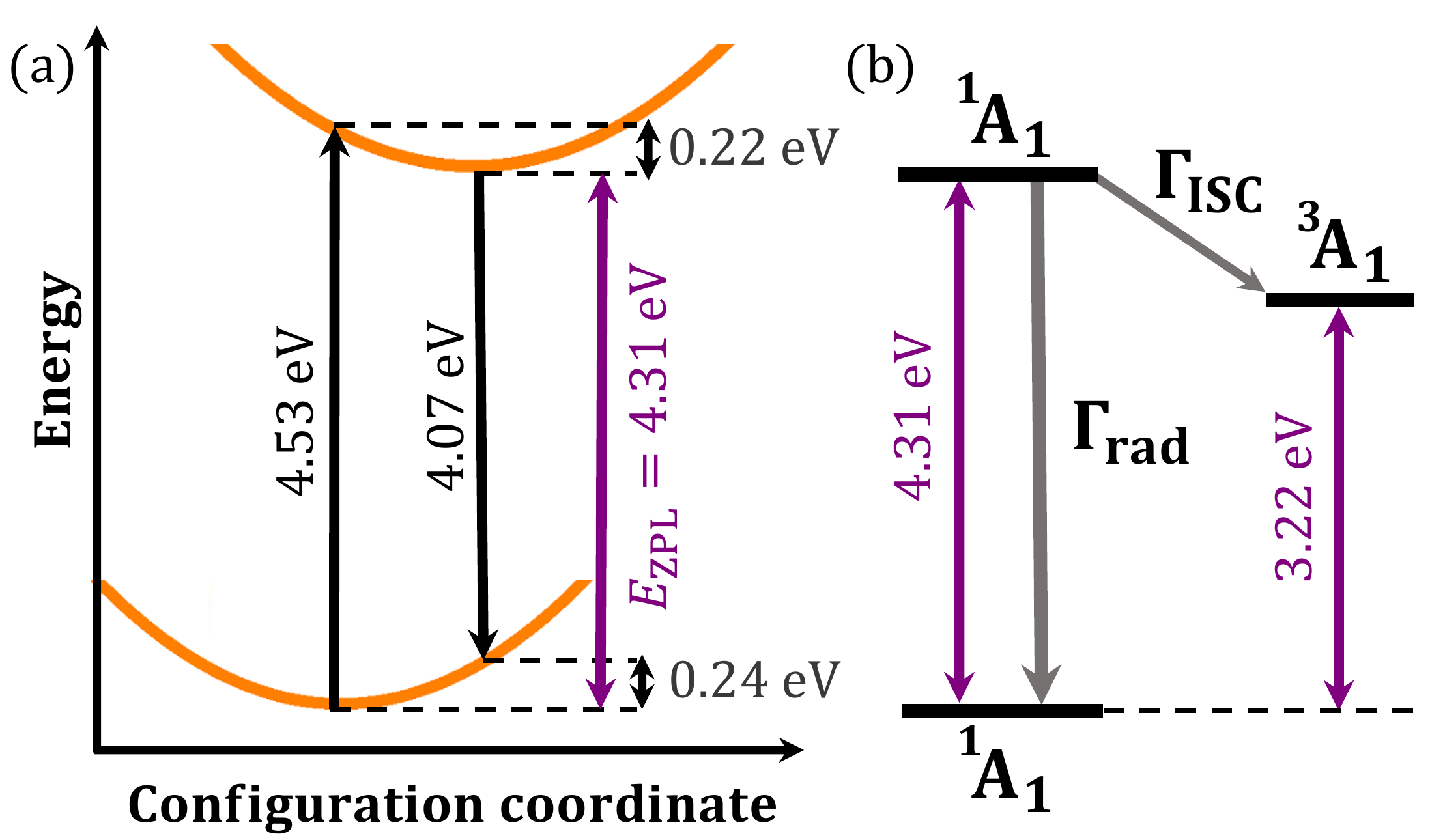}
\caption{(a) Configuration coordinate diagram describing the optical excitation process between two $^1A_1$ states for the neutral {$\CC$} dimer in hexagonal boron nitride.
(b) Energy-level diagram of the neutral {$\CC$} dimer.}
\label{fig:ccd}
\end{figure}

The calculated one-dimensional configuration coordinate diagram \cite{Stoneham} is shown in Fig.~\ref{fig:ccd}(a). We obtain a ZPL energy of $E_{\text{ZPL}}=4.31$ eV. The Franck-Condon shifts are $0.22$ eV in the excited state and $0.24$ eV in the ground state. To quantify electron-phonon coupling, we calculate \cite{Alkauskas2012} the Huang-Rhys factor $S$, which is a measure of the average number of phonons emitted during the optical transition\cite{Stoneham}. We find an effective phonon frequency of $\hbar\Omega=120$ meV, yielding the Huang-Rhys factor $S=0.24/0.12=2.0$, which is consistent with the experimental estimate $S=1-2$ reported in Ref.~\onlinecite{Vokhmintsev2019}. The Huang-Rhys factor is related to the Debye-Waller factor $w_{\text{ZPL}}$ (the fraction of light emitted into the ZPL) via $w_{\text{ZPL}}\approx e^{-S}$. Our calculated value of $w_{\text{ZPL}}\approx 0.14$ is smaller than the experimental value of $0.26$ reported in Ref.~\onlinecite{Vuong2016}. However, note that due to the exponential dependence of $w_{\text{ZPL}}$ on $S$ small errors in the latter can lead to large errors in the former. 

Apart from the excited-state singlet $^1A_1$, there is also a triplet state $^3A_1$ with configuration $\vert b_{2}{b}^*_{2}\rangle$. $^3A_1$ is 3.22 eV above the ground state, i.e., 1.09 eV lower than the excited-state singlet due to the exchange interaction between the two electrons. The energy-level diagram of the neutral {$\CC$} dimer is shown in Fig.~\ref{fig:ccd}(b).

The rate of the radiative transition between the two singlet states is given by (in SI units) \cite{Stoneham}:
\begin{equation}
\Gamma_{\text{rad}} =\frac{1}{\tau_{\text{rad}}}=\frac{n_D E_{\text{ZPL}}^{3} \mu^2} {3\pi\varepsilon_0 c^3 \hbar^4}.
\label{eq:rate}
\end{equation}
Here $\varepsilon_{0}$ is vacuum permittivity, $n_{D}$ is the refractive index of the host ($n_D\approx 2.6$ for energy $E\approx 4$ eV \cite{Cappellini2001}), and $\mu=1.06$ $e\text{\AA}$ is the computed transition dipole moment for the transition $b_2\rightarrow b_2^*$. Using the calculated value of $E_{\text{ZPL}}=4.31$ eV, we obtain the rate $\Gamma_{\text{rad}}=8.6\times10^{8}$ s$^{-1}$, corresponding to $\tau_{\text{rad}}=1.2$ ns.

The $b_2 \rightarrow b_2^*$ transition is a strong dipole transition (so-called  $\pi \rightarrow \pi^*$ transition) with the polarization along the C--C bond. Experimental measurements of polarization would be really valuable but have not yet been performed so far. The calculated $\tau_{\text{rad}}=1.2$ ns is in good agreement with the experimental value of $1.1-1.2$ ns \cite{Era1981,Museur2008}, but one should exercise caution comparing the two. The lifetime of the excited state $^1A_1$ is governed by two decay mechanisms [Fig.~\ref{fig:ccd}(b)]: the radiative transition to the ground state $\Gamma_{\text{rad}}$ and the inter-system crossing (ISC) to the triplet state $\Gamma_{\text{ISC}}$: $\tau=1/(\Gamma_\text{rad}+\Gamma_\text{ISC})$.  However, in the Supplementary Material we show that $\Gamma_{\text{ISC}}\ll \Gamma_{\text{rad}}$, and this justifies the comparison of the calculated rate with the measured one. Since $\Gamma_{\text{ISC}}\ll \Gamma_{\text{rad}}$, we also conclude that the quantum efficiency of the radiative transition is close to unity.

Our results show that the calculated optical properties of the {$\CC$} defect are in very good agreement with the known properties of the 4.1 eV line. In fact, carbon dimers {\it have} been observed by annular dark field (ADF) electron microscopy in boron nitride monolayers \cite{Krivanek2010} exfoliated from {\it bulk} hBN. Carbon atoms have distinct intensity in ADF images, and this allowed a direct identification of C--C pairs  \cite{Krivanek2010}. This experimental proof of the existence of {$\CC$} defects in bulk hBN inos in excellent agreement with our conclusions regarding the stability of carbon dimers.

In summary, we have reported the results of hybrid functional calculations for the carbon dimer in hexagonal boron nitride. Those calculations allow us to conclude that {$\CC$} is the defect which is responsible for the 4.1 eV emission in hBN. The carbon dimer is expected to form whenever carbon is present during growth, explaining the observed correlation between the presence of carbon and the 4.1 eV line. The calculated zero-phonon line of the intra-defect optical transition of $4.31$ eV is close to the experimental value. Moreover, the theoretical Huang-Rhys factor of $S=2.0$ is consistent with the experimental estimate $S = 1-2$, and radiative lifetime $\tau_{\text{rad}}=1.2$ ns is close to experimental value $\tau_{\text{rad}}=1.1 - 1.2$ ns. Identification of the chemical nature of the defect will enable more controlled experiments involving the 4.1 eV line, in particular using the {$\CC$} defect as a single photon emitter \cite{Bourrellier2016}. Our analysis shows that the quantum efficiency of this emitter should be close to unity. Combined with a very short radiative lifetime this results in a very high photon yield. Together with a modest value of the Huang-Rhys factor (large weight of the ZPL) and a well-defined polarization axis, this makes the carbon dimer a very interesting quantum emitter in the near UV.

\section*{Supplementary Material}
See Supplementary Material for the calculation of the excited state singlet and the inter-system crossing rate from the excited-state singlet $^1A_1$ to the $^3A_1$ state.

\section*{Acknowledgments}
MMS and MM contributed equally to this work.
We acknowledge useful discussions with Marcus W. Doherty, Lukas Razinkovas, and Mark Turiansky.
AA was funded by the Grant No.~9.3.3.-LMT-K-712-14-0085 from the Research Council of Lithuania.
MMS acknowledges funding from the European Union's Horizon 2020 research and innovation programme under grant agreement No. 820394 (project A{\sc steriqs}).
The research reported here was also partially supported by the National Science Foundation (NSF) through the Materials Research Science and Engineering Center at UC Santa Barbara, DMR-1720256 (Seed).
Calculations were performed at the High Performance Computing Center ``HPC Saul\.etekis'' in the Faculty of Physics, Vilnius University.
Computational resources were also provided by the Extreme Science and Engineering Discovery Environment (XSEDE), supported by the NSF (ACI-1548562).


%




\begin{thebibliography}{32}%
\makeatletter
\providecommand \@ifxundefined [1]{%
 \@ifx{#1\undefined}
}%
\providecommand \@ifnum [1]{%
 \ifnum #1\expandafter \@firstoftwo
 \else \expandafter \@secondoftwo
 \fi
}%
\providecommand \@ifx [1]{%
 \ifx #1\expandafter \@firstoftwo
 \else \expandafter \@secondoftwo
 \fi
}%
\providecommand \natexlab [1]{#1}%
\providecommand \enquote  [1]{``#1''}%
\providecommand \bibnamefont  [1]{#1}%
\providecommand \bibfnamefont [1]{#1}%
\providecommand \citenamefont [1]{#1}%
\providecommand \href@noop [0]{\@secondoftwo}%
\providecommand \href [0]{\begingroup \@sanitize@url \@href}%
\providecommand \@href[1]{\@@startlink{#1}\@@href}%
\providecommand \@@href[1]{\endgroup#1\@@endlink}%
\providecommand \@sanitize@url [0]{\catcode `\\12\catcode `\$12\catcode
  `\&12\catcode `\#12\catcode `\^12\catcode `\_12\catcode `\%12\relax}%
\providecommand \@@startlink[1]{}%
\providecommand \@@endlink[0]{}%
\providecommand \url  [0]{\begingroup\@sanitize@url \@url }%
\providecommand \@url [1]{\endgroup\@href {#1}{\urlprefix }}%
\providecommand \urlprefix  [0]{URL }%
\providecommand \Eprint [0]{\href }%
\providecommand \doibase [0]{http://dx.doi.org/}%
\providecommand \selectlanguage [0]{\@gobble}%
\providecommand \bibinfo  [0]{\@secondoftwo}%
\providecommand \bibfield  [0]{\@secondoftwo}%
\providecommand \translation [1]{[#1]}%
\providecommand \BibitemOpen [0]{}%
\providecommand \bibitemStop [0]{}%
\providecommand \bibitemNoStop [0]{.\EOS\space}%
\providecommand \EOS [0]{\spacefactor3000\relax}%
\providecommand \BibitemShut  [1]{\csname bibitem#1\endcsname}%
\let\auto@bib@innerbib\@empty
\bibitem [{\citenamefont {Duong}, \citenamefont {Yun},\ and\ \citenamefont
  {Lee}(2017)}]{Duong2017}%
  \BibitemOpen
  \bibfield  {author} {\bibinfo {author} {\bibfnamefont {D.~L.}\ \bibnamefont
  {Duong}}, \bibinfo {author} {\bibfnamefont {S.~J.}\ \bibnamefont {Yun}}, \
  and\ \bibinfo {author} {\bibfnamefont {Y.~H.}\ \bibnamefont {Lee}},\ }\href
  {\doibase 10.1021/acsnano.7b07436} {\bibfield  {journal} {\bibinfo  {journal}
  {ACS Nano}\ }\textbf {\bibinfo {volume} {11}},\ \bibinfo {pages} {11803}
  (\bibinfo {year} {2017})}\BibitemShut {NoStop}%
\bibitem [{\citenamefont {Cassabois}, \citenamefont {Valvin},\ and\
  \citenamefont {Gil}(2016)}]{Cassabois2016}%
  \BibitemOpen
  \bibfield  {author} {\bibinfo {author} {\bibfnamefont {G.}~\bibnamefont
  {Cassabois}}, \bibinfo {author} {\bibfnamefont {P.}~\bibnamefont {Valvin}}, \
  and\ \bibinfo {author} {\bibfnamefont {B.}~\bibnamefont {Gil}},\ }\href
  {\doibase 10.1038/nphoton.2015.277} {\bibfield  {journal} {\bibinfo
  {journal} {Nature Photonics}\ }\textbf {\bibinfo {volume} {10}},\ \bibinfo
  {pages} {262} (\bibinfo {year} {2016})}\BibitemShut {NoStop}%
\bibitem [{\citenamefont {Watanabe}, \citenamefont {Taniguchi},\ and\
  \citenamefont {Kanda}(2004)}]{Watanabe2004}%
  \BibitemOpen
  \bibfield  {author} {\bibinfo {author} {\bibfnamefont {K.}~\bibnamefont
  {Watanabe}}, \bibinfo {author} {\bibfnamefont {T.}~\bibnamefont {Taniguchi}},
  \ and\ \bibinfo {author} {\bibfnamefont {H.}~\bibnamefont {Kanda}},\ }\href
  {\doibase 10.1038/nmat1134} {\bibfield  {journal} {\bibinfo  {journal}
  {Nature Materials}\ }\textbf {\bibinfo {volume} {3}},\ \bibinfo {pages} {404}
  (\bibinfo {year} {2004})}\BibitemShut {NoStop}%
\bibitem [{\citenamefont {Kim}\ \emph {et~al.}(2018)\citenamefont {Kim},
  \citenamefont {Fr{\"o}ch}, \citenamefont {Christian}, \citenamefont {Straw},
  \citenamefont {Bishop}, \citenamefont {Totonjian}, \citenamefont {Watanabe},
  \citenamefont {Taniguchi}, \citenamefont {Toth},\ and\ \citenamefont
  {Aharonovich}}]{Kim2018}%
  \BibitemOpen
  \bibfield  {author} {\bibinfo {author} {\bibfnamefont {S.}~\bibnamefont
  {Kim}}, \bibinfo {author} {\bibfnamefont {J.~E.}\ \bibnamefont {Fr{\"o}ch}},
  \bibinfo {author} {\bibfnamefont {J.}~\bibnamefont {Christian}}, \bibinfo
  {author} {\bibfnamefont {M.}~\bibnamefont {Straw}}, \bibinfo {author}
  {\bibfnamefont {J.}~\bibnamefont {Bishop}}, \bibinfo {author} {\bibfnamefont
  {D.}~\bibnamefont {Totonjian}}, \bibinfo {author} {\bibfnamefont
  {K.}~\bibnamefont {Watanabe}}, \bibinfo {author} {\bibfnamefont
  {T.}~\bibnamefont {Taniguchi}}, \bibinfo {author} {\bibfnamefont
  {M.}~\bibnamefont {Toth}}, \ and\ \bibinfo {author} {\bibfnamefont
  {I.}~\bibnamefont {Aharonovich}},\ }\href {\doibase
  10.1038/s41467-018-05117-4} {\bibfield  {journal} {\bibinfo  {journal}
  {Nature Communications}\ }\textbf {\bibinfo {volume} {9}},\ \bibinfo {pages}
  {2623} (\bibinfo {year} {2018})}\BibitemShut {NoStop}%
\bibitem [{\citenamefont {Tran}\ \emph {et~al.}(2016)\citenamefont {Tran},
  \citenamefont {Bray}, \citenamefont {Ford}, \citenamefont {Toth},\ and\
  \citenamefont {Aharonovich}}]{Tran2016}%
  \BibitemOpen
  \bibfield  {author} {\bibinfo {author} {\bibfnamefont {T.~T.}\ \bibnamefont
  {Tran}}, \bibinfo {author} {\bibfnamefont {K.}~\bibnamefont {Bray}}, \bibinfo
  {author} {\bibfnamefont {M.~J.}\ \bibnamefont {Ford}}, \bibinfo {author}
  {\bibfnamefont {M.}~\bibnamefont {Toth}}, \ and\ \bibinfo {author}
  {\bibfnamefont {I.}~\bibnamefont {Aharonovich}},\ }\href {\doibase
  10.1038/nnano.2015.242} {\bibfield  {journal} {\bibinfo  {journal} {Nature
  Nanotechnology}\ }\textbf {\bibinfo {volume} {11}},\ \bibinfo {pages} {37}
  (\bibinfo {year} {2016})}\BibitemShut {NoStop}%
\bibitem [{\citenamefont {Larach}\ and\ \citenamefont
  {Shrader}(1956)}]{Larach1956}%
  \BibitemOpen
  \bibfield  {author} {\bibinfo {author} {\bibfnamefont {S.}~\bibnamefont
  {Larach}}\ and\ \bibinfo {author} {\bibfnamefont {R.~E.}\ \bibnamefont
  {Shrader}},\ }\href {\doibase 10.1103/PhysRev.104.68} {\bibfield  {journal}
  {\bibinfo  {journal} {Physical Review}\ }\textbf {\bibinfo {volume} {104}},\
  \bibinfo {pages} {68} (\bibinfo {year} {1956})}\BibitemShut {NoStop}%
\bibitem [{\citenamefont {Solozhenko}\ \emph {et~al.}(2001)\citenamefont
  {Solozhenko}, \citenamefont {Lazarenko}, \citenamefont {Petitet},\ and\
  \citenamefont {Kanaev}}]{Solozhenko2001}%
  \BibitemOpen
  \bibfield  {author} {\bibinfo {author} {\bibfnamefont {V.}~\bibnamefont
  {Solozhenko}}, \bibinfo {author} {\bibfnamefont {A.}~\bibnamefont
  {Lazarenko}}, \bibinfo {author} {\bibfnamefont {J.-P.}\ \bibnamefont
  {Petitet}}, \ and\ \bibinfo {author} {\bibfnamefont {A.}~\bibnamefont
  {Kanaev}},\ }\href {\doibase 10.1016/S0022-3697(01)00030-0} {\bibfield
  {journal} {\bibinfo  {journal} {Journal of Physics and Chemistry of Solids}\
  }\textbf {\bibinfo {volume} {62}},\ \bibinfo {pages} {1331} (\bibinfo {year}
  {2001})}\BibitemShut {NoStop}%
\bibitem [{\citenamefont {Era}, \citenamefont {Minami},\ and\ \citenamefont
  {Kuzuba}(1981)}]{Era1981}%
  \BibitemOpen
  \bibfield  {author} {\bibinfo {author} {\bibfnamefont {K.}~\bibnamefont
  {Era}}, \bibinfo {author} {\bibfnamefont {F.}~\bibnamefont {Minami}}, \ and\
  \bibinfo {author} {\bibfnamefont {T.}~\bibnamefont {Kuzuba}},\ }\href
  {\doibase 10.1016/0022-2313(81)90223-4} {\bibfield  {journal} {\bibinfo
  {journal} {Journal of Luminescence}\ }\textbf {\bibinfo {volume} {24}},\
  \bibinfo {pages} {71} (\bibinfo {year} {1981})}\BibitemShut {NoStop}%
\bibitem [{\citenamefont {Museur}, \citenamefont {Feldbach},\ and\
  \citenamefont {Kanaev}(2008)}]{Museur2008}%
  \BibitemOpen
  \bibfield  {author} {\bibinfo {author} {\bibfnamefont {L.}~\bibnamefont
  {Museur}}, \bibinfo {author} {\bibfnamefont {E.}~\bibnamefont {Feldbach}}, \
  and\ \bibinfo {author} {\bibfnamefont {A.}~\bibnamefont {Kanaev}},\ }\href
  {\doibase 10.1103/PhysRevB.78.155204} {\bibfield  {journal} {\bibinfo
  {journal} {Physical Review B}\ }\textbf {\bibinfo {volume} {78}},\ \bibinfo
  {pages} {155204} (\bibinfo {year} {2008})}\BibitemShut {NoStop}%
\bibitem [{\citenamefont {Du}\ \emph {et~al.}(2015)\citenamefont {Du},
  \citenamefont {Li}, \citenamefont {Lin},\ and\ \citenamefont
  {Jiang}}]{Du2015}%
  \BibitemOpen
  \bibfield  {author} {\bibinfo {author} {\bibfnamefont {X.}~\bibnamefont
  {Du}}, \bibinfo {author} {\bibfnamefont {J.}~\bibnamefont {Li}}, \bibinfo
  {author} {\bibfnamefont {J.}~\bibnamefont {Lin}}, \ and\ \bibinfo {author}
  {\bibfnamefont {H.}~\bibnamefont {Jiang}},\ }\href {\doibase
  10.1063/1.4905908} {\bibfield  {journal} {\bibinfo  {journal} {Applied
  Physics Letters}\ }\textbf {\bibinfo {volume} {106}},\ \bibinfo {pages}
  {021110} (\bibinfo {year} {2015})}\BibitemShut {NoStop}%
\bibitem [{\citenamefont {Stoneham}(2001)}]{Stoneham}%
  \BibitemOpen
  \bibfield  {author} {\bibinfo {author} {\bibfnamefont {A.~M.}\ \bibnamefont
  {Stoneham}},\ }\href@noop {} {\emph {\bibinfo {title} {Theory of defects in
  solids}}}\ (\bibinfo  {publisher} {Oxford University Press},\ \bibinfo {year}
  {2001})\BibitemShut {NoStop}%
\bibitem [{\citenamefont {Vokhmintsev}, \citenamefont {Weinstein},\ and\
  \citenamefont {Zamyatin}(2019)}]{Vokhmintsev2019}%
  \BibitemOpen
  \bibfield  {author} {\bibinfo {author} {\bibfnamefont {A.}~\bibnamefont
  {Vokhmintsev}}, \bibinfo {author} {\bibfnamefont {I.}~\bibnamefont
  {Weinstein}}, \ and\ \bibinfo {author} {\bibfnamefont {D.}~\bibnamefont
  {Zamyatin}},\ }\href {\doibase 10.1016/j.jlumin.2018.12.036} {\bibfield
  {journal} {\bibinfo  {journal} {Journal of Luminescence}\ }\textbf {\bibinfo
  {volume} {208}},\ \bibinfo {pages} {363} (\bibinfo {year}
  {2019})}\BibitemShut {NoStop}%
\bibitem [{\citenamefont {Bourrellier}\ \emph {et~al.}(2016)\citenamefont
  {Bourrellier}, \citenamefont {Meuret}, \citenamefont {Tararan}, \citenamefont
  {St{\'e}phan}, \citenamefont {Kociak}, \citenamefont {Tizei},\ and\
  \citenamefont {Zobelli}}]{Bourrellier2016}%
  \BibitemOpen
  \bibfield  {author} {\bibinfo {author} {\bibfnamefont {R.}~\bibnamefont
  {Bourrellier}}, \bibinfo {author} {\bibfnamefont {S.}~\bibnamefont {Meuret}},
  \bibinfo {author} {\bibfnamefont {A.}~\bibnamefont {Tararan}}, \bibinfo
  {author} {\bibfnamefont {O.}~\bibnamefont {St{\'e}phan}}, \bibinfo {author}
  {\bibfnamefont {M.}~\bibnamefont {Kociak}}, \bibinfo {author} {\bibfnamefont
  {L.~H.}\ \bibnamefont {Tizei}}, \ and\ \bibinfo {author} {\bibfnamefont
  {A.}~\bibnamefont {Zobelli}},\ }\href {\doibase 10.1021/acs.nanolett.6b01368}
  {\bibfield  {journal} {\bibinfo  {journal} {Nano Letters}\ }\textbf {\bibinfo
  {volume} {16}},\ \bibinfo {pages} {4317} (\bibinfo {year}
  {2016})}\BibitemShut {NoStop}%
\bibitem [{\citenamefont {Meuret}\ \emph {et~al.}(2015)\citenamefont {Meuret},
  \citenamefont {Tizei}, \citenamefont {Cazimajou}, \citenamefont
  {Bourrellier}, \citenamefont {Chang}, \citenamefont {Treussart},\ and\
  \citenamefont {Kociak}}]{Mueret2015}%
  \BibitemOpen
  \bibfield  {author} {\bibinfo {author} {\bibfnamefont {S.}~\bibnamefont
  {Meuret}}, \bibinfo {author} {\bibfnamefont {L.~H.~G.}\ \bibnamefont
  {Tizei}}, \bibinfo {author} {\bibfnamefont {T.}~\bibnamefont {Cazimajou}},
  \bibinfo {author} {\bibfnamefont {R.}~\bibnamefont {Bourrellier}}, \bibinfo
  {author} {\bibfnamefont {H.~C.}\ \bibnamefont {Chang}}, \bibinfo {author}
  {\bibfnamefont {F.}~\bibnamefont {Treussart}}, \ and\ \bibinfo {author}
  {\bibfnamefont {M.}~\bibnamefont {Kociak}},\ }\href {\doibase
  10.1103/PhysRevLett.114.197401} {\bibfield  {journal} {\bibinfo  {journal}
  {Physical Review Letters}\ }\textbf {\bibinfo {volume} {114}},\ \bibinfo
  {pages} {197401} (\bibinfo {year} {2015})}\BibitemShut {NoStop}%
\bibitem [{\citenamefont {Uddin}\ \emph {et~al.}(2017)\citenamefont {Uddin},
  \citenamefont {Li}, \citenamefont {Lin},\ and\ \citenamefont
  {Jiang}}]{Uddin2017}%
  \BibitemOpen
  \bibfield  {author} {\bibinfo {author} {\bibfnamefont {M.}~\bibnamefont
  {Uddin}}, \bibinfo {author} {\bibfnamefont {J.}~\bibnamefont {Li}}, \bibinfo
  {author} {\bibfnamefont {J.}~\bibnamefont {Lin}}, \ and\ \bibinfo {author}
  {\bibfnamefont {H.}~\bibnamefont {Jiang}},\ }\href {\doibase
  10.1063/1.4982647} {\bibfield  {journal} {\bibinfo  {journal} {Applied
  Physics Letters}\ }\textbf {\bibinfo {volume} {110}},\ \bibinfo {pages}
  {182107} (\bibinfo {year} {2017})}\BibitemShut {NoStop}%
\bibitem [{\citenamefont {Katzir}\ \emph {et~al.}(1975)\citenamefont {Katzir},
  \citenamefont {Suss}, \citenamefont {Zunger},\ and\ \citenamefont
  {Halperin}}]{Katzir1975}%
  \BibitemOpen
  \bibfield  {author} {\bibinfo {author} {\bibfnamefont {A.}~\bibnamefont
  {Katzir}}, \bibinfo {author} {\bibfnamefont {J.}~\bibnamefont {Suss}},
  \bibinfo {author} {\bibfnamefont {A.}~\bibnamefont {Zunger}}, \ and\ \bibinfo
  {author} {\bibfnamefont {A.}~\bibnamefont {Halperin}},\ }\href {\doibase
  10.1103/PhysRevB.11.2370} {\bibfield  {journal} {\bibinfo  {journal}
  {Physical Review B}\ }\textbf {\bibinfo {volume} {11}},\ \bibinfo {pages}
  {2370} (\bibinfo {year} {1975})}\BibitemShut {NoStop}%
\bibitem [{\citenamefont {Weston}\ \emph {et~al.}(2018)\citenamefont {Weston},
  \citenamefont {Wickramaratne}, \citenamefont {Mackoit}, \citenamefont
  {Alkauskas},\ and\ \citenamefont {Van~de Walle}}]{Weston2018}%
  \BibitemOpen
  \bibfield  {author} {\bibinfo {author} {\bibfnamefont {L.}~\bibnamefont
  {Weston}}, \bibinfo {author} {\bibfnamefont {D.}~\bibnamefont
  {Wickramaratne}}, \bibinfo {author} {\bibfnamefont {M.}~\bibnamefont
  {Mackoit}}, \bibinfo {author} {\bibfnamefont {A.}~\bibnamefont {Alkauskas}},
  \ and\ \bibinfo {author} {\bibfnamefont {C.~G.}\ \bibnamefont {Van~de
  Walle}},\ }\href {\doibase 10.1103/PhysRevB.97.214104} {\bibfield  {journal}
  {\bibinfo  {journal} {Physical Review B}\ }\textbf {\bibinfo {volume} {97}},\
  \bibinfo {pages} {214104} (\bibinfo {year} {2018})}\BibitemShut {NoStop}%
\bibitem [{\citenamefont {Korona}\ and\ \citenamefont
  {Chojecki}(2019)}]{Korona2019}%
  \BibitemOpen
  \bibfield  {author} {\bibinfo {author} {\bibfnamefont {T.}~\bibnamefont
  {Korona}}\ and\ \bibinfo {author} {\bibfnamefont {M.}~\bibnamefont
  {Chojecki}},\ }\href {\doibase 10.1002/qua.25925} {\bibfield  {journal}
  {\bibinfo  {journal} {International Journal of Quantum Chemistry}\ }\textbf
  {\bibinfo {volume} {119}},\ \bibinfo {pages} {e25925} (\bibinfo {year}
  {2019})}\BibitemShut {NoStop}%
\bibitem [{\citenamefont {Heyd}, \citenamefont {Scuseria},\ and\ \citenamefont
  {Ernzerhof}(2003)}]{HSE}%
  \BibitemOpen
  \bibfield  {author} {\bibinfo {author} {\bibfnamefont {J.}~\bibnamefont
  {Heyd}}, \bibinfo {author} {\bibfnamefont {G.~E.}\ \bibnamefont {Scuseria}},
  \ and\ \bibinfo {author} {\bibfnamefont {M.}~\bibnamefont {Ernzerhof}},\
  }\href {\doibase 10.1063/1.1564060} {\bibfield  {journal} {\bibinfo
  {journal} {Journal of Chemical Physics}\ }\textbf {\bibinfo {volume} {118}},\
  \bibinfo {pages} {8207} (\bibinfo {year} {2003})}\BibitemShut {NoStop}%
\bibitem [{\citenamefont {Perdew}, \citenamefont {Burke},\ and\ \citenamefont
  {Ernzerhof}(1996)}]{PBE}%
  \BibitemOpen
  \bibfield  {author} {\bibinfo {author} {\bibfnamefont {J.~P.}\ \bibnamefont
  {Perdew}}, \bibinfo {author} {\bibfnamefont {K.}~\bibnamefont {Burke}}, \
  and\ \bibinfo {author} {\bibfnamefont {M.}~\bibnamefont {Ernzerhof}},\ }\href
  {\doibase 10.1103/PhysRevLett.77.3865} {\bibfield  {journal} {\bibinfo
  {journal} {Physical Review Letters}\ }\textbf {\bibinfo {volume} {77}},\
  \bibinfo {pages} {3865} (\bibinfo {year} {1996})}\BibitemShut {NoStop}%
\bibitem [{\citenamefont {Antonius}\ \emph {et~al.}(2015)\citenamefont
  {Antonius}, \citenamefont {Ponc\'e}, \citenamefont {Lantagne-Hurtubise},
  \citenamefont {Au\-clair}, \citenamefont {Gonze},\ and\ \citenamefont
  {C\^ot\'e}}]{Antonius2015}%
  \BibitemOpen
  \bibfield  {author} {\bibinfo {author} {\bibfnamefont {G.}~\bibnamefont
  {Antonius}}, \bibinfo {author} {\bibfnamefont {S.}~\bibnamefont {Ponc\'e}},
  \bibinfo {author} {\bibfnamefont {E.}~\bibnamefont {Lantagne-Hurtubise}},
  \bibinfo {author} {\bibfnamefont {G.}~\bibnamefont {Au\-clair}}, \bibinfo
  {author} {\bibfnamefont {X.}~\bibnamefont {Gonze}}, \ and\ \bibinfo {author}
  {\bibfnamefont {M.}~\bibnamefont {C\^ot\'e}},\ }\href {\doibase
  10.1103/PhysRevB.92.085137} {\bibfield  {journal} {\bibinfo  {journal}
  {Physical Review B}\ }\textbf {\bibinfo {volume} {92}},\ \bibinfo {pages}
  {085137} (\bibinfo {year} {2015})}\BibitemShut {NoStop}%
\bibitem [{\citenamefont {Bl{\"o}chl}(1994)}]{PAW}%
  \BibitemOpen
  \bibfield  {author} {\bibinfo {author} {\bibfnamefont {P.~E.}\ \bibnamefont
  {Bl{\"o}chl}},\ }\href {\doibase 10.1103/PhysRevB.50.17953} {\bibfield
  {journal} {\bibinfo  {journal} {Physical Review B}\ }\textbf {\bibinfo
  {volume} {50}},\ \bibinfo {pages} {17953} (\bibinfo {year}
  {1994})}\BibitemShut {NoStop}%
\bibitem [{\citenamefont {Grimme}(2006)}]{Grimme2006}%
  \BibitemOpen
  \bibfield  {author} {\bibinfo {author} {\bibfnamefont {S.}~\bibnamefont
  {Grimme}},\ }\href {\doibase 10.1002/jcc.20495} {\bibfield  {journal}
  {\bibinfo  {journal} {Journal of Computational Chemistry}\ }\textbf {\bibinfo
  {volume} {27}},\ \bibinfo {pages} {1787} (\bibinfo {year}
  {2006})}\BibitemShut {NoStop}%
\bibitem [{\citenamefont {Tomaszkiewicz}(2002)}]{Tomaszkiewicz2002}%
  \BibitemOpen
  \bibfield  {author} {\bibinfo {author} {\bibfnamefont {I.}~\bibnamefont
  {Tomaszkiewicz}},\ }\href
  {http://www.ichf.edu.pl/pjch/pj-2002/pj-2002-06.htm} {\bibfield  {journal}
  {\bibinfo  {journal} {Polish Journal of Chemistry}\ }\textbf {\bibinfo
  {volume} {76}},\ \bibinfo {pages} {891} (\bibinfo {year} {2002})}\BibitemShut
  {NoStop}%
\bibitem [{\citenamefont {Kresse}\ and\ \citenamefont
  {Furthm{\"u}ller}(1996)}]{vasp}%
  \BibitemOpen
  \bibfield  {author} {\bibinfo {author} {\bibfnamefont {G.}~\bibnamefont
  {Kresse}}\ and\ \bibinfo {author} {\bibfnamefont {J.}~\bibnamefont
  {Furthm{\"u}ller}},\ }\href {\doibase 10.1103/PhysRevB.54.11169} {\bibfield
  {journal} {\bibinfo  {journal} {Physical Review B}\ }\textbf {\bibinfo
  {volume} {54}},\ \bibinfo {pages} {11169} (\bibinfo {year}
  {1996})}\BibitemShut {NoStop}%
\bibitem [{\citenamefont {Freysoldt}\ \emph {et~al.}(2014)\citenamefont
  {Freysoldt}, \citenamefont {Grabowski}, \citenamefont {Hickel}, \citenamefont
  {Neugebauer}, \citenamefont {Kresse}, \citenamefont {Janotti},\ and\
  \citenamefont {Van~de Walle}}]{Freysoldt2014}%
  \BibitemOpen
  \bibfield  {author} {\bibinfo {author} {\bibfnamefont {C.}~\bibnamefont
  {Freysoldt}}, \bibinfo {author} {\bibfnamefont {B.}~\bibnamefont
  {Grabowski}}, \bibinfo {author} {\bibfnamefont {T.}~\bibnamefont {Hickel}},
  \bibinfo {author} {\bibfnamefont {J.}~\bibnamefont {Neugebauer}}, \bibinfo
  {author} {\bibfnamefont {G.}~\bibnamefont {Kresse}}, \bibinfo {author}
  {\bibfnamefont {A.}~\bibnamefont {Janotti}}, \ and\ \bibinfo {author}
  {\bibfnamefont {C.~G.}\ \bibnamefont {Van~de Walle}},\ }\href {\doibase
  10.1103/RevModPhys.86.253} {\bibfield  {journal} {\bibinfo  {journal}
  {Reviews of Modern Physics}\ }\textbf {\bibinfo {volume} {86}},\ \bibinfo
  {pages} {253} (\bibinfo {year} {2014})}\BibitemShut {NoStop}%
\bibitem [{\citenamefont {Freysoldt}, \citenamefont {Neugebauer},\ and\
  \citenamefont {Van~de Walle}(2009)}]{Freysoldt2009}%
  \BibitemOpen
  \bibfield  {author} {\bibinfo {author} {\bibfnamefont {C.}~\bibnamefont
  {Freysoldt}}, \bibinfo {author} {\bibfnamefont {J.}~\bibnamefont
  {Neugebauer}}, \ and\ \bibinfo {author} {\bibfnamefont {C.~G.}\ \bibnamefont
  {Van~de Walle}},\ }\href {\doibase 10.1103/PhysRevLett.102.016402} {\bibfield
   {journal} {\bibinfo  {journal} {Physical Review Letters}\ }\textbf {\bibinfo
  {volume} {102}},\ \bibinfo {pages} {016402} (\bibinfo {year}
  {2009})}\BibitemShut {NoStop}%
\bibitem [{\citenamefont {Jones}\ and\ \citenamefont
  {Gunnarsson}(1989)}]{Jones1989}%
  \BibitemOpen
  \bibfield  {author} {\bibinfo {author} {\bibfnamefont {R.~O.}\ \bibnamefont
  {Jones}}\ and\ \bibinfo {author} {\bibfnamefont {O.}~\bibnamefont
  {Gunnarsson}},\ }\href {\doibase 10.1103/RevModPhys.61.689} {\bibfield
  {journal} {\bibinfo  {journal} {Reviews of Modern Physics}\ }\textbf
  {\bibinfo {volume} {61}},\ \bibinfo {pages} {689} (\bibinfo {year}
  {1989})}\BibitemShut {NoStop}%
\bibitem [{\citenamefont {Alkauskas}\ \emph {et~al.}(2012)\citenamefont
  {Alkauskas}, \citenamefont {Lyons}, \citenamefont {Steiauf},\ and\
  \citenamefont {Van~de Walle}}]{Alkauskas2012}%
  \BibitemOpen
  \bibfield  {author} {\bibinfo {author} {\bibfnamefont {A.}~\bibnamefont
  {Alkauskas}}, \bibinfo {author} {\bibfnamefont {J.~L.}\ \bibnamefont
  {Lyons}}, \bibinfo {author} {\bibfnamefont {D.}~\bibnamefont {Steiauf}}, \
  and\ \bibinfo {author} {\bibfnamefont {C.~G.}\ \bibnamefont {Van~de Walle}},\
  }\href {\doibase 10.1103/PhysRevLett.109.267401} {\bibfield  {journal}
  {\bibinfo  {journal} {Physical Review Letters}\ }\textbf {\bibinfo {volume}
  {109}},\ \bibinfo {pages} {267401} (\bibinfo {year} {2012})}\BibitemShut
  {NoStop}%
\bibitem [{\citenamefont {Vuong}\ \emph {et~al.}(2016)\citenamefont {Vuong},
  \citenamefont {Cassabois}, \citenamefont {Valvin}, \citenamefont {Ouerghi},
  \citenamefont {Chassagneux}, \citenamefont {Voisin},\ and\ \citenamefont
  {Gil}}]{Vuong2016}%
  \BibitemOpen
  \bibfield  {author} {\bibinfo {author} {\bibfnamefont {T.}~\bibnamefont
  {Vuong}}, \bibinfo {author} {\bibfnamefont {G.}~\bibnamefont {Cassabois}},
  \bibinfo {author} {\bibfnamefont {P.}~\bibnamefont {Valvin}}, \bibinfo
  {author} {\bibfnamefont {A.}~\bibnamefont {Ouerghi}}, \bibinfo {author}
  {\bibfnamefont {Y.}~\bibnamefont {Chassagneux}}, \bibinfo {author}
  {\bibfnamefont {C.}~\bibnamefont {Voisin}}, \ and\ \bibinfo {author}
  {\bibfnamefont {B.}~\bibnamefont {Gil}},\ }\href@noop {} {\bibfield
  {journal} {\bibinfo  {journal} {Physical Review Letters}\ }\textbf {\bibinfo
  {volume} {117}},\ \bibinfo {pages} {097402} (\bibinfo {year}
  {2016})}\BibitemShut {NoStop}%
\bibitem [{\citenamefont {Cappellini}\ \emph {et~al.}(2001)\citenamefont
  {Cappellini}, \citenamefont {Satta}, \citenamefont {Palummo},\ and\
  \citenamefont {Onida}}]{Cappellini2001}%
  \BibitemOpen
  \bibfield  {author} {\bibinfo {author} {\bibfnamefont {G.}~\bibnamefont
  {Cappellini}}, \bibinfo {author} {\bibfnamefont {G.}~\bibnamefont {Satta}},
  \bibinfo {author} {\bibfnamefont {M.}~\bibnamefont {Palummo}}, \ and\
  \bibinfo {author} {\bibfnamefont {G.}~\bibnamefont {Onida}},\ }\href
  {\doibase 10.1103/PhysRevB.64.035104} {\bibfield  {journal} {\bibinfo
  {journal} {Physical Review B}\ }\textbf {\bibinfo {volume} {64}},\ \bibinfo
  {pages} {035104} (\bibinfo {year} {2001})}\BibitemShut {NoStop}%
\bibitem [{\citenamefont {Krivanek}\ \emph {et~al.}(2010)\citenamefont
  {Krivanek}, \citenamefont {Chisholm}, \citenamefont {Nicolosi}, \citenamefont
  {Pennycook}, \citenamefont {Corbin}, \citenamefont {Dellby}, \citenamefont
  {Murfitt}, \citenamefont {Own}, \citenamefont {Szilagyi}, \citenamefont
  {Oxley}, \citenamefont {Pantelides},\ and\ \citenamefont
  {Pennycook}}]{Krivanek2010}%
  \BibitemOpen
  \bibfield  {author} {\bibinfo {author} {\bibfnamefont {O.~L.}\ \bibnamefont
  {Krivanek}}, \bibinfo {author} {\bibfnamefont {M.~F.}\ \bibnamefont
  {Chisholm}}, \bibinfo {author} {\bibfnamefont {V.}~\bibnamefont {Nicolosi}},
  \bibinfo {author} {\bibfnamefont {T.~J.}\ \bibnamefont {Pennycook}}, \bibinfo
  {author} {\bibfnamefont {G.~J.}\ \bibnamefont {Corbin}}, \bibinfo {author}
  {\bibfnamefont {N.}~\bibnamefont {Dellby}}, \bibinfo {author} {\bibfnamefont
  {M.~F.}\ \bibnamefont {Murfitt}}, \bibinfo {author} {\bibfnamefont {C.~S.}\
  \bibnamefont {Own}}, \bibinfo {author} {\bibfnamefont {Z.~S.}\ \bibnamefont
  {Szilagyi}}, \bibinfo {author} {\bibfnamefont {M.~P.}\ \bibnamefont {Oxley}},
  \bibinfo {author} {\bibfnamefont {S.~T.}\ \bibnamefont {Pantelides}}, \ and\
  \bibinfo {author} {\bibfnamefont {S.~J.}\ \bibnamefont {Pennycook}},\ }\href
  {\doibase 10.1038/nature08879} {\bibfield  {journal} {\bibinfo  {journal}
  {Nature}\ }\textbf {\bibinfo {volume} {464}},\ \bibinfo {pages} {571}
  (\bibinfo {year} {2010})}\BibitemShut {NoStop}%
\end{thebibliography}
\end{document}


\title{Carbon dimer defect as a source of the 4.1 eV luminescence in hexagonal boron nitride: Supplementary material}

\author{Ma\v{z}ena Mackoit-Sinkevi\v{c}ien\.{e}}
\author{Marek Maciaszek}
\author{Chris G. Van de Walle}
\author{Audrius Alkauskas}
\noaffiliation
\maketitle


\section{Calculation of the excited state singlet}

In the bra-ket notation the excited state singlet is given by:
\begin{gather}
\left | S \right \rangle = \frac{1}{\sqrt{2}} \left ( \left | b_2 \bar{b}_2^*\right \rangle - \left | \bar{b}_2 b_{2}^* \right \rangle \right ).
\end{gather}
The three components of the triplet state are:
\begin{gather}
\left | T; -1 \right \rangle =  \left | \bar{b}_2 \bar{b}_2^* \right \rangle 
\nonumber \\
\left | T; 0 \right \rangle = \frac{1}{\sqrt{2}} \left ( \left | b_2 \bar{b}_2^*\right \rangle + \left | \bar{b}_2 b_{2}^* \right \rangle \right )
\\
\left | T; +1 \right \rangle =  \left | b_2 b_2^* \right \rangle 
\nonumber
\end{gather}
Let us evaluate the expectation value of the Hamiltonian $\mathcal{H}$ in the mixed-spin state $\left | S/T \right \rangle \equiv \left |  b_2 \bar{b}_2^* \right \rangle = \frac{1}{\sqrt{2}} \left ( \left | S \right \rangle + \left | T; 0 \right \rangle \right )$:
\begin{gather}
 \left \langle b_2 \bar{b}_2^*   \left | \mathcal{H}  \right | b_2 \bar{b}_2^* \right \rangle 
= \frac{1}{2} \left (  
  \left \langle S  \left | \mathcal{H}  \right | S \right \rangle 
+ \left \langle T; 0  \left | \mathcal{H}  \right | T; 0 \right \rangle 
\right ).
\label{mixed}
\end{gather}
The cross term $\left \langle S  \left | \mathcal{H}  \right | T;0 \right \rangle $ disappears because the singlet state and the triplet state are both eigenfunctions of $\mathcal{H}$ and therefore orthogonal. Using the fact that triplet states are degenerate, from Eq.~\eqref{mixed} we find that $E(S) \equiv \left \langle S  \left | \mathcal{H}  \right | S \right \rangle  =  2\left \langle b_2 \bar{b}_2^*   \left | \mathcal{H}  \right | b_2 \bar{b}_2^* \right \rangle - \left \langle T; + 1 \left | \mathcal{H}  \right | T; +1 \right \rangle$, or:
\begin{gather}
 E(S) = 2E(S/T) - E(T).
\end{gather}
This is our final expression for the energy of the excited state singlet. Here $E(S/T)$ and $E(T)$ are the energies of the mixed-spin state $\left | b_2 \bar{b}_2^* \right \rangle$ and the triplet state  $\left |  b_2 b_2^* \right \rangle $, respectively. Since these two states are composed of a single Slater determinant, their energies can be calculated using the $\Delta$SCF method by promoting one electron from the orbital $b_2$ to the orbital $b_2^*$. In the calculations we assume that equilibrium geometries of the mixed-spin state and the excited state singlet are very similar, since they correspond to the same nominal electron configuration.

\section{Inter-system crossing}

Here we provide the estimate for the inter-system crossing (ISC) rate $\Gamma_{\text{ISC}}$ between excited states $^1A_1$ and $^3A_1$ of the $\CC$ defect [Figure 3(b) of the main text]. The principal mechanisms for inter-system crossing are \cite{Penfold2018} (i) spin-orbit interaction and (ii) hyperfine interaction.  

(i) Spin-orbit interaction can be written in the form \cite{Tinkham} $V_{so}=\sum_{i=1,2}\vec{\lambda}_i\cdot\vec{\sigma}_i$, where the sum is over the two optically active electrons of the $\CC$ defect, $\vec{\sigma}$ is a vector made of Pauli matrices, and $\vec{\lambda}$ quantifies the spin-orbit interaction (mean-field approximation is assumed). In the $C_{2v}$ point group the three Cartesian components of $\vec{\lambda}$ transform like $B_2$, $B_1$, and $A_2$ irreducible representations \cite{Tinkham}, and therefore they do not couple two states with $A_1$ orbital symmetry. Thus, we conclude that to first order spin-orbit interactions will cause no ISC between $^1A_1$ and $^3A_1$ states. 

(ii) To estimate the contribution of hyperfine interactions we first note that both defect orbitals are localized on carbon atoms [Figure 2(b) of the main text]. Since the majority of carbon nuclei are $^{12}$C with no nuclear moment, only hyperfine interactions with more distant B and N nuclei will contribute. The ISC rate due to hyperfine interactions rate is given by the Fermi golden rule $\Gamma_{hf}=(2\pi/\hbar) V_{hf}^2L(\Delta E)$, where $V_{hf}$ is hyperfine coupling strength, and $L(\Delta E)$ is a lineshape function for the transition $^1A_1 \rightarrow ^3$$A_1$, identical to the lineshape functions in optical transitions \cite{Alkauskas2012} ($\Delta E=1.09$ eV is the energy difference between the two states, cf.~Figure 3 of the main text). In Ref.~\cite{Exarhos2019} it was estimated that hyperfine coupling of electrons in $p_z$-type orbitals with B or N nuclei that reside {\it on the site} where the orbital is localized is on the order of 100 MHz. Even if we assume the same coupling constant with more distant nuclei, we obtain rates much smaller than 1 Hz showing that ISC due to hyperfine coupling is negligible. 

This analysis allows us to conclude that $\Gamma_{\text{ISC}}\ll \Gamma_{\text{rad}}$ (see main text), and therefore the decay of the excited state $^1A_1$ will be mainly due to radiative decay. This justifies the comparison of the calculated radiative rate with the measured rate of luminescence decay.